\documentclass[aps,prl,twocolumn,groupedaddress,amsmath,amssymb]{revtex4-1}
\usepackage{graphicx,subfigure,bm}

\begin{document}


\title{Spectral solution of urn models for interacting particle systems}


\author{William Pickering}
\author{Chjan Lim}
\affiliation{Department of Mathematical Sciences, Rensselaer Polytechnic Institute, 110 8th Street, Troy, New York 12180, USA}

\date{\today}
\begin{abstract}
Using generating function methods for diagonalizing the transition
matrix in 2-Urn models, we provide a complete classification into
solvable and unsolvable subclasses, with further division of the solvable
models into the Martingale and non-Martingale subcategories, and
prove that the stationary distribution is a Gaussian function in the
latter. We also give a natural condition related to the symmetry
of the random walk in which the non-Martingale Urn models lead to an increase in entropy from Gaussian states. The condition also shows that universal symmetry in the macro-state is equivalent to increasing entropy. Certain models of social opinion
dynamics, treated as Urn models, do not increase in entropy,
unlike isolated mechanical systems.

\end{abstract}

\pacs{}

\maketitle

Physical applications of urn problems can be traced to the Ehrenfest model to describe the Second Law of Thermodynamics \cite{ehrenfest}. However, systems such as the Ehrenfest model do not adequately describe the dynamics of interacting particle systems \cite{liggett2}. In the systems that we introduce, the particles change urns by interacting with each other. Naturally, these models have a wide range of physical applications for various interpretations of the urns themselves, such as well mixed kinetic reactions \cite{henriksen,upadhyay,frachebourg} and thermodynamics \cite{ehrenfest,kac}. These models also have applications to social opinion dynamics, in which the voter model  \cite{liggett,clifford,castellano} is the only case that is a martingale. When the system is generalized to three urns, one can pose Naming Game dynamics on the complete graph \cite{baronchelli2, xie,zhang,sen} in a similar fashion. Further instances of interacting particle systems are the contact process, exclusion processes, and stochastic Ising models \cite{liggett,liggett2}.

\begin{figure}[h!]
\includegraphics[scale=.4]{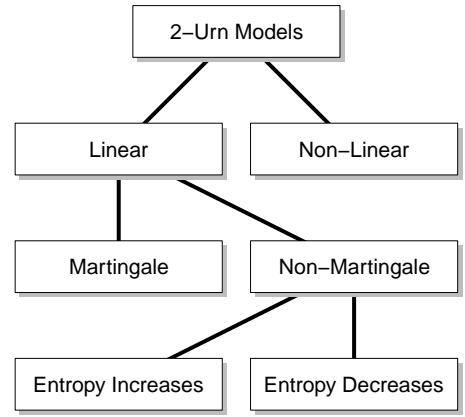}
\caption{Classification tree for the 2-Urn problems that shows all relevant subclasses. Among the linear cases $(2\gamma_1-2\gamma_2-\alpha_1+\alpha_2+\beta_1-\beta_2=0)$, the martingales $(\alpha_1=\beta_1, \alpha_2=\beta_2=\gamma_1=\gamma_2=0)$ are equivalent to the voter model and the non-martingales constitute a much larger class of models. We show that an entropy increase in the non-martingale cases is equivalent to macroscopic symmetry of the model given by Eqn. \eqref{macro_sym}.}
\label{fig1}
\end{figure}

In addition to the class of models that describe interacting particle systems, we also provide their exact solutions. The method is an extension of the generating function solution of the Ehrenfest model formulated by Mark Kac in 1947 \cite{kac}. We utilize a generating function method for solving the spectral problem of the Markov transition matrix for each model. With the explicit diagonalization of the transition matrix, we can compute several quantities depending on the application of the model. In sociophysics \cite{sen}, the expected time to consensus is one quantity of interest \cite{vazquez,zhang,sood,pickering,cooper,baronchelli2}. We also provide a condition in which entropy will decrease from Gaussian states, which would violate the Second Law of Thermodynamics. Fig. \ref{fig1} describes the classification of these models based upon solvability and relevant macroscopic properties.

In these models, two urns ($A$ and $B$) have $N$ balls distributed between them. In a discrete time step, two balls are drawn randomly. The balls are redistributed between the urns stochastically. The redistribution probabilities depend on the urns from which the balls came and the order that they were drawn.

The post-selection probability distributions define rate parameters which specify the model in the macro-state. Let $n_A(m)$ denote the number of balls in urn $A$ at discrete time $m$. The model is characterized through the specification of the rate parameters, which we denote as $\{\alpha_1,\alpha_2,\beta_1,\beta_2,\gamma_1,\gamma_2\}$. Given that the balls came from different urns, $\alpha_1$ and $\beta_1$ are the probabilities that $n_A$ increases/decreases by one respectively. If they were drawn from the same urn, $\alpha_2$ and $\beta_2$ are the probabilities that $n_A$ increases or decreases by one respectively. Similarly, $\gamma_1$ and $\gamma_2$ are the probabilities that $n_A$ increases or decreases by two. Since these parameters correspond to probabilities, they are constrained as such, e.g., $\alpha_2+\gamma_1\leq1$. 

The parameters of the urn model correspond to the influence of people in social settings. Values of $\alpha_1$ and $\beta_1$ correspond to the impact a person has on another with the opposite opinion. Here, two opposing individuals enter a discussion and one of them changes their opinion as a result. The voter model assumes this and has parameter configuration $\{\alpha,0,\alpha,0,0,0\}$, where $\alpha$ scales time. The other parameters, $\alpha_2,\beta_2,\gamma_1,$ and $\gamma_2$ can correspond to mutation and competition between individuals. Also, these parameters can represent push-pull factors to Lee's model of migration \cite{lee}, and quadratic transition probabilities reflect the assumptions made in Gravity models of migration and trade \cite{lewer,anderson}. Existing models with explicit parameter configurations also include the Moran model of genetic drift, with parameters $\{1-\mu_1,\mu_2,1-\mu_2,\mu_1,0,0\}$, where $\mu_1$ and $\mu_2$ are mutation probabilities \cite{moran,blythe}. We can explicitly diagonalize this model for any $N$ and choice of parameters using these techniques. These models emphasize that the population size, $N$, is not always large and thus the discrete analysis here is necessary. We can also solve an extension with $\gamma_1,\gamma_2\not=0$ that is beyond the tridiagonal case, which the Stieltjes integral representation method of Karlin cannot solve \cite{karlin1,karlin2}. 

The parameters affect the transition probabilities of the urn model when $n_A=i$, which are given to be
\begin{align}
p_i^{(1)}&=\alpha_1\frac{i(N-i)}{N(N-1)}+\alpha_2\frac{(N-i)(N-i-1)}{N(N-1)}\\
p_i^{(2)}&=\gamma_1\frac{(N-i)(N-i-1)}{N(N-1)}\\
q_i^{(1)}&=\beta_1\frac{i(N-i)}{N(N-1)}+\beta_2\frac{i(i-1)}{N(N-1)}\\
q_i^{(2)}&=\gamma_2\frac{i(i-1)}{N(N-1)}.
\end{align}
Here, we define $p_i^{(k)}=Pr\{\Delta n_A=k|n_A=i\}$ and $q_i^{(k)}=Pr\{\Delta n_A=-k|n_A=i\}$. Notice that the parameter choice \{1,1,1,1,0,0\} exactly simplifies to the Ehrenfest model. Let $a_i^{(m)}=Pr\{n_A(m)=i\}$. We introduce the finite difference operator $\Delta_{ki}$ acting on a grid function $\phi_i$ defined as $\Delta_{ki}[\phi_i]=\phi_{i+k}-\phi_i$. We form the single step difference equation that describes the probability distribution in macro-state:
\begin{multline}
a_i^{(m+1)}-a_i^{(m)}=\Delta_{-1i}[p_i^{(1)}a_i^{(m)}]+\Delta_{-2i}[p_i^{(2)}a_i^{(m)}]\\
+\Delta_{+1i}[q_i^{(1)}a_i^{(m)}]+\Delta_{+2i}[q_i^{(2)}a_i^{(m)}].\label{singlepropagator}
\end{multline}
This constitutes a pentadiagonal Markov transition matrix for the system. We solve for all eigenvalues and eigenvectors of this model by extending the procedure in \cite{pickering}. For eigenvalue $\lambda$ and eigenvector $\mathbf{v}$ with components $c_i$, let $G(x,y)=\sum_i c_ix^iy^{N-i}$ be the generating function for the eigenvectors. We rewrite the spectral problem for the single step propagator given in Eqn. \eqref{singlepropagator} as a partial differential equation for $G$ using the differentiation and shift properties of $G$ \cite{newman,newman2,bender,pickering}. The PDE for $G$ is

\begin{multline}
N(N-1)(\lambda-1)G=\gamma_1(x^2-y^2)G_{yy}+\alpha_1x(x-y)G_{xy}\\
+\alpha_2y(x-y)G_{yy}-\gamma_2(x^2-y^2)G_{xx}-\beta_1y(x-y)G_{xy}\\
-\beta_2x(x-y)G_{xx}.
\end{multline}

To solve this equation, we make the change of variables $u=x-y$ and $H(u,y)=G(x,y)$. We show below that $H$ has the same structure as $G$. That is, we define $H(u,y)=\sum_i b_iu^iy^{N-i}$. This change of variables allows us to solve the system when $2\gamma_1-2\gamma_2-\alpha_1+\alpha_2+\beta_1-\beta_2=0$. Under this restriction, the change of variables will transform the pentadiagonal structure of the transition matrix into a lower triangular matrix. Since the transformed matrix is lower triangular, the difference equation for $b_i$ is explicit, which allows us to find both $\lambda$ and $c_i$. Collecting coefficients in the transformed PDE for $H$ yields

\begin{widetext}
\begin{equation}
b_i=\frac{\{[(-2\gamma_1+\alpha_1)(i-1)+(2\gamma_1+\alpha_2)(N-i)]b_{i-1}+\gamma_1(N-i+2)b_{i-2}\}(N-i+1)}{N(N-1)(\lambda-1)+(2\gamma_1+2\gamma_2+\alpha_2+\beta_2)i(N-i)+\frac{1}{2}(\alpha_1+\alpha_2+\beta_1+\beta_2)i(i-1)}.\label{b}
\end{equation}
\end{widetext}

This allows us to find all eigenvalues exactly. Since $c_i=0$  for $i<0$ and $i>N$, we require $b_i=0$ for $i<0$ and $i>N$ as well. Since Eqn. \eqref{b} is an explicit linear difference equation, every $b_i=0$ unless the equation is singular for some $i=k$. However, this corresponds to the trivial solution to the eigenvalue problem. Thus, the denominator of Eq. \eqref{b} must be zero when $i=k$. Solving for $\lambda$ shows that the eigenvalues are

\begin{multline}
\lambda_k=1-\frac{(2\gamma_1+2\gamma_2+\alpha_2+\beta_2)k(N-k)}{N(N-1)}\\
-\frac{(\alpha_1+\alpha_2+\beta_1+\beta_2)k(k-1)}{2N(N-1)}
\end{multline}
for $k=0\ldots N$. This allows $b_k$ to take any value. Values for $b_i$ for any $i>k$ can be found by repeated application of Eq. \eqref{b}. Expressing $H(u,y)$ in the original coordinates, gives
\begin{equation}
G(x,y)=\sum_{i=0}^N\sum_{j=i}^N (-1)^{j-i}{j\choose i}b_j x^iy^{N-i},
\end{equation}
which shows that $H$ and $G$ have the same form \cite{pickering}. Thus the spectral problem is solved for all urn models that satisfy $2\gamma_1-2\gamma_2-\alpha_1+\alpha_2+\beta_1-\beta_2=0$. This parameter constraint holds if and only if the dynamical system for mean density $[\bar{\rho}_A(t),\bar{\rho}_B(t)]^T$ is linear. We define a \textit{linear urn model} as any of the above models that satisfy this constraint. We conjecture that no other change of variables $(x,y)\rightarrow (u,v)$ will solve the nonlinear cases in this fashion, although a proof of this claim is not given.

The treatment of the spectral problem by generating functions is equivalent to a similarity transformation of the transition matrix. Let $\mathbf{T}$ denote the transition matrix given by Eqn. \eqref{singlepropagator} and let $\mathbf{v}=\mathbf{Pw}$ for some transformation matrix $\mathbf{P}$. Then, the spectral problem for $\mathbf{w}$ is given by $\mathbf{P}^{-1}\mathbf{TPw}=\lambda\mathbf{w}$. The generating function method prescribes the matrix $\mathbf{P}$ so that the new matrix $\mathbf{L}=\mathbf{P}^{-1}\mathbf{TP}$ is lower triangular with a bandwidth of at most two. The components of the transformation matrices that do this are determined to be
\begin{align}
&[\mathbf{P}]_{ij}=(-1)^{j-i} {j\choose i}\\
&[\mathbf{P}^{-1}]_{ij}={j\choose i}.
\end{align}
We use the convention that ${j\choose i}=0$ when $i>j$, which suggest that $\mathbf{P}$ and $\mathbf{P}^{-1}$ are upper triangular. 

The spectral decomposition of the transition matrix can be found by this similarity transformation. We do this by diagonalizing the matrix $\mathbf{L}=\mathbf{W\Lambda W}^{-1}$. Here, $\mathbf{\Lambda}=diag(\lambda_0,\ldots ,\lambda_N)$ and $\mathbf{W}$ are the eigenvectors of $\mathbf{L}$. The components of these eigenvectors are $b_i$ corresponding to eigenvalue $\lambda_k$. Since $b_i=0$ for $i<j$, $\mathbf{W}$ is lower triangular. Therefore, $\mathbf{W}^{-1}$ can be found explicitly via forward substitution. Diagonalization of $\mathbf{L}$ allows us to explicitly diagonalize the transition matrix as
\begin{equation}
\mathbf{T}=\mathbf{(PW)\Lambda}\mathbf{(PW)}^{-1}\label{diagonalization}.
\end{equation}

The immediate consequence of the explicit diagonalization of the transition matrix is the solution of the $m$ step propagator. With this solution, we can find several valuable quantities summarized in Table \ref{solutions}. The quantity $d_k$ is the initial distribution expressed in the eigenbasis. That is, $d_k$ are the components of $\mathbf{d}=\mathbf{(PW)}^{-1}\mathbf{a}^{(0)}$.

\begin{table}
\caption{Exact Solutions}\label{solutions}
\begin{tabular}{p{2cm} c}
\hline
\hline
Quantity & Discrete Solution\\
\hline

Macro-state probability & $a_i^{(m)}=\sum_{k=0}^N d_k[\mathbf{v}_k]_i\lambda_k^m$\\

Consensus time & $\displaystyle E[\tau^p]\sim\sum_{k=1}^N\frac{d_kp!}{[N(1-\lambda_k)]^{p+1}}\times$\\
&$\displaystyle \bigg\{\alpha_1[\mathbf{v}_k]_{N-1}+\frac{2\gamma_1}{N-1}[\mathbf{v}_k]_{N-2}\bigg\}$\\
&\\
Local time & $E[\mathbf{M}]\sim\displaystyle\frac{1}{N}\mathop{\sum_{k=0}^N}_{\lambda_k\not=1}\frac{d_k}{1-\lambda_k}\mathbf{v}_k$\\
&\\
Gibbs entropy & $\displaystyle S(m)=-\sum_{i=0}^N a_i^{(m)}\log\frac{a_i^{(m)}}{{N \choose i}}$\\
&\\
\hline
\hline
\end{tabular}
\end{table}

\begin{figure}[h!]
\includegraphics[scale=.4]{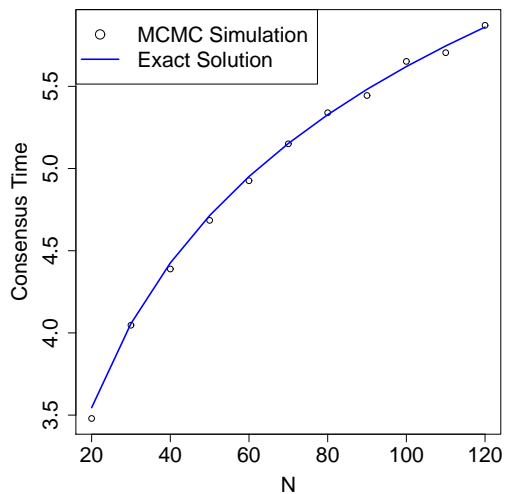}
\caption{Consensus time simulation data plotted with the exact solution given in Table \ref{solutions}. For each $N$, the expected time to consensus is averaged over 1000 runs of the urn model with parameters \{1, 1/4, 1/4, 0, 1/4, 0\}. Using drift estimation \cite{baronchelli2,castello}, it can be shown that the consensus time is $O(\log N)$.}
\label{fig2}
\end{figure}

The consensus time is the amount of scaled time $(\tau=m/N)$ until all of the balls are in a single urn and the dynamics of the system halt. We assume that only one of the consensus points is an absorbing state and without loss in generality, we assume that it is when $n_A=N$ instead of $n_A=0$. When both consensus points are absorbing, the linear urn model reduces to the voter model on the complete graph, which is well studied \cite{zhang,pickering,sood, yildiz}. Not only can the expected time to consensus be found, but the $p^{th}$ moment as well \cite{pickering}. Fig. \ref{fig2} compares the exact solution against simulation data.

The local time is the total amount of scaled time spent in each macro-state prior to the absorbing consensus. The sum of the local times is equal to the consensus time, which makes this a more detailed quantity. The expected local time $M_i$ for macro-state $n_A=i$ is known to be $E[M_i]=\frac{1}{N}\sum_{m=0}^\infty a_i^{(m)}$, which we can compute exactly by the diagonalization \cite{pickering}.

Next, we consider entropy defined in the sense of Gibbs \cite{jaynes}. We also make the assumption that each micro-state is equally likely for a given macro-state \cite{laurendeau}. When the probability distribution is $\mathcal{N}(\mu,\sigma^2)$, the entropy can be shown to be $S\sim N[\log2-2\sigma^2-2(\mu-\frac{1}{2})^2]+\log(2\sqrt{N}\sigma)+\frac{1}{2}$. We use this when calculating the change in entropy between the stationary distribution and an initial Gaussian distribution, $\mathcal{N}(\mu_0,\sigma_0^2)$. Let $\pi_i$ to be the discrete stationary distribution and let $\pi(\rho ; N)\approx \pi_iN$ approximate $\pi_i$ for large $N$. The following results show that the stationary distribution is asymptotically Gaussian whose mean and variance are also calculated exactly.

\textbf{Theorem: }\textit{The following holds for all non-martingale linear urn models $[2\gamma_1+2\gamma_2+\alpha_2+\beta_2\not=0]$:}
\begin{align}
&(1)\;\; \pi(\rho;N)= \mathcal{N}(\mu_f,\sigma_f^2)+O\left(\frac{1}{\sqrt{N}}\right)\\
&(2)\;\; E_\pi[n_A]=b_1\\
&(3)\;\; Var_\pi(n_A)=2b_2+b_1-b_1^2
\end{align}
\textit{Proof:} The proof of statement (1) begins by finding the Fokker-Plank equation for the probability density, $u(\rho,t)$, as $N\rightarrow\infty$:
\begin{equation}
u_t\sim-\frac{\partial}{\partial\rho}[v(\rho)u]+\frac{1}{2N}\frac{\partial^2}{\partial\rho^2}[D(\rho)u].
\end{equation}
Here, $v(\rho)=2p^{(2)}(\rho)+p^{(1)}(\rho)-q^{(1)}(\rho)-2q^{(2)}(\rho)$, and $D(\rho)=4p^{(2)}(\rho)+p^{(1)}(\rho)+q^{(1)}(\rho)+4q^{(2)}(\rho)$. The functions $p^{(k)}(\rho)$ and $q^{(k)}(\rho)$ are the continuous analogs of their discrete counterparts. As $t\rightarrow\infty$, $u(t,\rho)\rightarrow\pi(\rho;N)$, and $u_t(\rho,t)\rightarrow0$. The ODE for the stationary distribution is
\begin{equation}
0=-\frac{d}{d\rho}[v(\rho)\pi(\rho;N)]+\frac{1}{2N}\frac{d^2}{d\rho^2}[D(\rho)\pi(\rho;N)]. \label{stat_ODE}
\end{equation}
Since we assume that the drift is linear, $v(\rho)$ is a linear function. Furthermore, there exists $\rho_0$ such that $v(\rho_0)=0$. Therefore, we write $v(\rho)=v_1(\rho-\rho_0)$. We make the change of variables $\rho\rightarrow\xi$ defined as $\rho=\rho_0+\delta\xi$. We will choose $\delta=o(1)$ so that the ODE has non-trivial balance. Also, we let $\pi(\rho;N)=\pi_0(\rho;N)+\delta\pi_1(\rho;N)+\ldots$ be an asymptotic expansion of the stationary probability density. Making these substitutions into Eqn. \eqref{stat_ODE} yields

\begin{equation}
-v_1\frac{d}{d\xi}[\xi\pi_0(\xi;N)]+\frac{D_0}{2N\delta^2}\frac{d^2}{d\xi^2}[\pi_0(\xi;N)]+O\left(\frac{1}{N\delta}\right)=0.
\end{equation}
The coefficients $D_0$ and $v_1$ depend only of the choice of parameters $\{\alpha_1,\alpha_2,\beta_1,\beta_2,\gamma_1,\gamma_2\}$. For the leading order terms to balance, choose $\delta=1/\sqrt{N}$. As $N\rightarrow\infty$, the leading order terms give an ODE for $\pi_0(\xi;N)$. The solution of this ODE is a Gaussian function centered at $\xi=0$. To characterize the mean and variance, we prove statements $(2)$ and $(3)$.

The proof of $(2)$ follows by considering the generating function $G(x,y)$ that corresponds to the eigenvalue $\lambda=1$. This is defined by $G(x,y)=\sum_i \pi_ix^iy^{N-i}$. Note that $G(1,1)=1$ for normalization and $E_\pi[n_A]=G_x(1,1)$. In terms of $H(u,y)=\sum_i b_iu^iy^{N-i}$, we have that $H(0,1)=1$ and $G_x(1,1)=H_u(0,1)$. This implies that, $b_0=1$ and $H_u(0,1)=b_1$. Therefore, $E_\pi[n_A]=b_1$. Since $b_i=0$ for $i<0$, we can use Eqn. \eqref{b} to find $b_1$ exactly. The proof of $(3)$ is demonstrated in a similar fashion.  $\square$

The values of $b_1$ and $b_2$ depend only on the choice of parameters and $N$, which allow us to exactly characterize the stationary distribution. When $b_2$ and $b_1$ are combined to give $Var_\pi(n_A)$, the result is $O(N)$. Thus, the variance for the density is $\sigma_f^2=O(1/N)$. Since the stationary distribution is also Gaussian, the change in entropy is $\Delta S\sim\log(\sigma_f/\sigma_0)-2N\sigma^2_f+2N[\sigma_0^2+(\mu_0-1/2)^2-(\mu_f-1/2)^2]$. As $N\rightarrow\infty$, this implies that the entropy of the system will decrease when

\begin{equation}
\mu_f(1-\mu_f)<\mu_0(1-\mu_0)-\sigma_0^2. \label{entropy_condition}
\end{equation}

A significant consequence of Eqn. \eqref{entropy_condition} is that unless $\mu_f(1-\mu_f)$ achieves its maximum value, there will always exist an initial condition, $(\mu_0,\sigma_0^2)$, that will cause entropy to decrease with time for large $N$. Thus, for an entropy increase as $N\rightarrow\infty$, we require $\mu_f=b_1/N=1/2$.

We show that the condition given in Eqn. \eqref{entropy_condition} for an increase in entropy is equivalent to a form of symmetry in the model. Let $U_1$, $U_2$ denote the urns that the first and second balls are selected respectively and $U^c$ denotes the urn that is opposite to $U$. We define a \textit{macroscopically symmetric} urn model when 
\begin{equation}
E[\Delta n_A|U_1,U_2]=E[\Delta n_B|U_1^c, U_2^c] \label{macro_sym}
\end{equation}
for all permutations of $U_1,U_2$. The following result relates entropy to this form of symmetry.

\textbf{Theorem: }\textit{If the urn model is linear and non-martingale, then Eqn. \eqref{macro_sym} is necessary and sufficient for an increase in entropy from Gaussian states.}\\

\textit{Proof:} If the system is macroscopically symmetric, then $2\gamma_1-2\gamma_2+\alpha_2-\beta_2=0$ and $\alpha_1=\beta_1$. These constraints together imply that the system has linear drift. Furthermore, using Eqn \eqref{b}, we have that $\mu_f=1/2$, which is sufficient to show that entropy increases from Gaussian initial states by Eqn. \eqref{entropy_condition}. To show the converse, an entropy increase requires $\mu_f=1/2$, which implies $2\gamma_1-2\gamma_2+\alpha_2-\beta_2=0$. Using linearity, we have $\alpha_1=\beta_1$, which is sufficient to show macroscopic symmetry. $\square$

If the microscopic behavior of the system is invariant under an urn permutation, then $\alpha_1=\beta_1,\alpha_2=\beta_2,\gamma_1=\gamma_2$. If microscopic symmetry holds, then macroscopic symmetry holds, but the converse is not necessarily true. A counterexample to this is to consider $\gamma_1=1/2, \beta_2=1$, and all other parameters are zero. 

The methods above may be useful in solving more sophisticated problems that involve more than two urns. For example, the multi-state voter model \cite{starnini}, the Naming Game \cite{waagen, baronchelli2}, and genetic drift with migration \cite{blythe} are multi-urn models with quadratic transition probabilities such as the above. The above techniques are capable of analyzing parts of these models to a greater extent and can provide more detailed solutions than existing methods.

\section*{Acknowledgement}

\begin{acknowledgments}
This work was supported in part by the Army Research
Office Grant No. W911NF-09-1-0254 and W911NF-12-
1- 467 0546. The views and conclusions contained in
this document are those of the authors and should not
be interpreted as representing the official policies, either
expressed or implied, of the Army Research Office or the
U.S. Government.
\end{acknowledgments}


\bibliography{2urnPRL1c}{}

\end{document}